\documentclass[twocolumn]{jpsj3}
\usepackage{txfonts}

\title{Muon Spin Relaxation and Neutron Diffraction Studies of Cluster-Glass States in Sr$_{1-x}$La$_{x}$RuO$_3$ }

\author{Ikuto~Kawasaki$^{1}$\thanks{E-mail address: kawasaki@sci.u-hyogo.ac.jp},  Kenji~Fujimura$^{2}$, Isao~Watanabe$^{3}$, Maxim~Avdeev$^{4}$, Kenichi~Tenya$^{5}$, and Makoto~Yokoyama$^{2}$  
}

\inst{\address{$^{1}$Graduate School of Material Science, University of Hyogo, Kamigori, Hyogo 678-1297, Japan} \\
\address{$^{2}$Faculty of Science, Ibaraki University, Mito, Ibaraki 310-8512, Japan} \\
\address{$^{3}$Advanced Meson Science Laboratory, RIKEN Nishina Center for Accelerator Based Science, RIKEN, Wako, Saitama 351-0198, Japan} \\
\address{$^{4}$Australian Nuclear Science and Technology Organization, Lucas Heights,  NSW 2234, Australia} \\
\address{$^{5}$Faculty of Education, Shinshu University, Nagano 390-8621, Japan} \\

}

\abst{To clarify the magnetic properties of cluster-glass states in Sr$_{1-x}$La$_{x}$RuO$_3$ (0.3 $\leq$ $x$ $\leq$ 0.5), we report herein the results of muon spin relaxation ($\mu$SR) and neutron powder diffraction measurements. The $\mu$SR experiments showed that magnetic clusters start developing well above the peak temperature $T^{*}$ in the ac susceptibility. The volume fraction of the magnetically ordered region  increases continuously with decreasing temperature, showing no anomaly at $T^{*}$, and  reaches nearly 100\% at the lowest temperature. The temperature variation of the volume fraction is essentially independent of the  La concentration in the $x$ range presently investigated, although the dc magnetization is significantly suppressed with increasing $x$.  Neutron powder diffraction experiments revealed that the ground state for $x$ = 0.3 is a long-range ferromagnetic ordered state. These results indicate that, with decreasing temperature, cluster-glass states in Sr$_{1-x}$La$_{x}$RuO$_3$ gradually develop into long-range ferromagnetic ordered states with decreasing temperature, and that the magnetic ordering process differs strikingly from that expected for a conventional second-order ferromagnetic transition.  }

\begin{document}
\maketitle

\section{Introduction}
In general, the 4$d$ orbitals in 4$d$ transition-metal oxides are more delocalized than the 3$d$ orbitals in 3$d$ transition-metal oxides; therefore, a conventional band picture is considered a suitable starting point for understanding such 4$d$ electronic states. However, 4$d$ transition-metal oxides exhibit a variety of intriguing electronic properties, such as quantum criticality,\cite{gri,gri2} non-Fermi liquid behavior,\cite{khali} and unconventional superconductivity.\cite{maeno}  These intriguing electronic properties imply that the correlation effect of 4$d$ electrons is important, and that the electronic states are not trivial.

Although 4$d$ transition-metal oxides rarely show magnetic ordering, SrRuO$_3$ exhibits ferromagnetic order below $T_\mathrm{C}$ = 160 K, and the ordered moment is approximately 1.1 $\mu_\mathrm{B}$.\cite{calla,kanba} This compound  has a nearly cubic structure of the GdFeO$_3$ type, which belongs to the space group $Pnma$.    The electronic structure has been studied by photoemission spectroscopy, which revealed that the valence band spectra, which consist of Ru 4$d$ and  O 2$p$ states, are roughly reproduced by band-structure calculations, and that the density of states at the Fermi level is dominated by Ru 4$d$ states.\cite{fujio_ph,okamoto_ph,park_ph,kim_ph,taki_ph,siemon_ph,grebin_ph} In addition, the size of the ordered moment is well reproduced by band-structure calculations.\cite{mazin}
These results suggest that the ferromagnetic order of this material is caused by itinerant Ru 4$d$ electrons and can be described within the context of the band-structure-based Stoner theory. 

However, an incoherent feature reflecting  electronic correlation effects is observed in the photoemission spectra. The Rhodes-Wohlfarth ratio $p_\mathrm{c}/p_\mathrm{s}$, where $p_\mathrm{c}$ and $p_\mathrm{s}$ are the paramagnetic dipole moment deduced from the Curie constant and the ordered ferromagnetic moment at zero temperature, respectively, is estimated to be approximately 1.3.\cite{fukunaga} This value is close to the localized-moment limit ($\sim$1).   The temperature dependence of the nuclear spin-relaxation rate deviates from the prediction of the self-consistent renormalization (SCR) theory for itinerant ferromagnets.\cite{yoshimura} Optical spectroscopy measurements have also shown that the charge dynamics differs from that predicted by Fermi liquid theory.\cite{kostic,doge2} Moreover, angle-resolved photoemission, magneto-optic Kerr effect, and optical spectroscopy experiments reveal that the ferromagnetic exchange splitting persists above  $T_\mathrm{C}$; thus, the presence of local moments  has been suggested.\cite{kerr,jeong,shai} 
These results reflect a localized character for the Ru 4$d$ states. Based on these results, it is considered that the itinerant and localized characters coexist in this compound; this coexistence is a fascinating aspect of this compound.

Solid solutions Sr$_{1-x}$La$_x$RuO$_3$ are of interest, since increases in the Ru-O distance and the RuO$_6$ octahedra rotation by La doping change the electronic states and may enhance the electronic correlation effect.\cite{mazin,nakatsu}  Recently, we have studied the electronic and magnetic properties of Sr$_{1-x}$La$_x$RuO$_3$ by means of macroscopic measurements of the specific heat, electrical resistivity,  ac susceptibility, and dc magnetization.\cite{kawasaki}  We showed that the ferromagnetic order is strongly suppressed with increasing La concentration $x$ and that the ordered state varies from ferromagnetic states to cluster-glass states for $x\geq0.3$, demonstrating that spatially inhomogeneous magnetic states, which are indicative of the localized character of Ru 4$d$ states, occur for $x\geq0.3$.  On the other hand, we also revealed that this system is metallic and that the density of states for conduction electrons at the  Fermi level is dominated by the Ru 4$d$ states over the entire range of La concentration investigated ($0\leq x\leq0.5$). 
Therefore, the itinerant and localized characters seem to coexist also in this solid solutions as in SrRuO$_3$.

On the other hand, microscopic details of cluster-glass states for $x\geq0.3$, which are very important for discussing the nature of the Ru 4$d$ states, still remain unclear. In the present study, we have performed muon spin relaxation ($\mu$SR) and neutron powder diffraction experiments on Sr$_{1-x}$La$_x$RuO$_3$ to characterize the cluster-glass states at the microscopic level. The  $\mu$SR and neutron powder diffraction techniques are complementary: the former technique provides information on local magnetic fields and magnetic fluctuations with extreme sensitivity and allows us to evaluate the volume fraction of  magnetic clusters.  In contrast, the latter technique provides information on the magnetic structure and correlation length.

\section{Experimental Details}

Polycrystalline samples of Sr$_{1-x}$La$_{x}$RuO$_3$ with $x$ = 0.3, 0.4, and 0.5 were prepared by the conventional solid-state reaction method.  Details of the sample preparation and characterization are given in Ref. 23. To estimate the ordering temperature and to characterize the magnetic properties of the present samples, we measured the ac susceptibility and dc magnetization using a commercial superconducting quantum interference device magnetometer in the temperature range of 2 to 300 K.    
The $\mu$SR experiments  were performed at the RIKEN-RAL Muon Facility in the UK, where an intense pulsed muon beam is available. 
The $\mu$SR spectra were collected under  zero-field (ZF), longitudinal-field (LF), and transverse-field (TF) conditions over a temperature range from 1.5 to 200 K in a He-flow cryostat. 
The residual magnetic field was reduced to less than 50 mOe by cancellation coils in the ZF-measurements.
The forward and backward counters are located upstream and downstream of the beam line, respectively. The muon spin asymmetry as a function of time is given by the expression $A(t)$ = $[N_+(t)$ $-$ $\alpha N_-(t)]/[N_+(t)$ $+$ $\alpha N_-(t)]$, where $N_+(t)$ and $N_-(t)$ represent the total muon events counted at time  $t$ by the forward and backward counters, respectively. The calibration factor $\alpha$ represents the relative counting efficiencies of the forward and backward counters and was estimated from a $\mu$SR measurement with an applied TF of 20 Oe. The samples were glued onto a high-purity silver plate and were tightly covered with a  25-$\mu$m-thick high-purity silver foil  to achieve good thermal contact. Any muons implanted onto the silver plate contribute to the $\mu$SR spectra in the form of a nearly time-independent constant background in the ZF- and LF-experiments.    
 
Neutron powder diffraction experiments were conducted for $x$ = 0.3 using the powder diffractometer Echidna at the OPAL facility of the Australian Nuclear Science and Technology Organisation.\cite{echidna}  The wavelength was set to 2.4395 \AA \ using a Ge (331) monochromator. Diffraction patterns were recorded over the angular range $20^\circ < 2\theta < 164^\circ$ and the temperature range of 1.4-50 K.  Approximately 10 g of the powdered sample were contained in a 9-mm-diameter cylindrical vanadium can, and the sample temperature was controlled by a closed-cycle He cryostat.

\section{Results}

\begin{figure}[h]
\begin{center}
\includegraphics[keepaspectratio, width=7cm,bb = 0 0 256 220,clip]{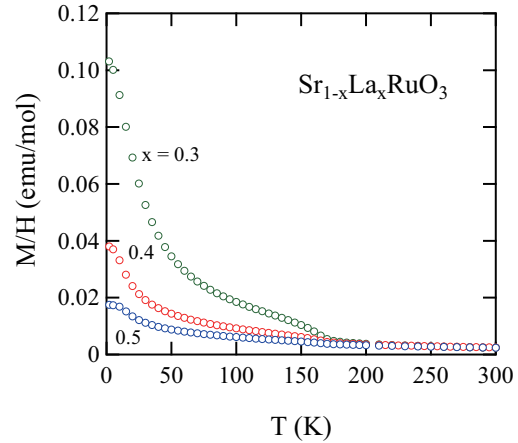}
\end{center}
\caption{(Color online) Temperature dependence of the dc magnetization for Sr$_{1-x}$La$_{x}$RuO$_3$ measured under a field of 5 kOe and under field-cooled conditions. }
\label{f1}
\end{figure}

\begin{figure}[h]
\begin{center}
\includegraphics[keepaspectratio, width=7cm,bb = 0 0 256 220,clip]{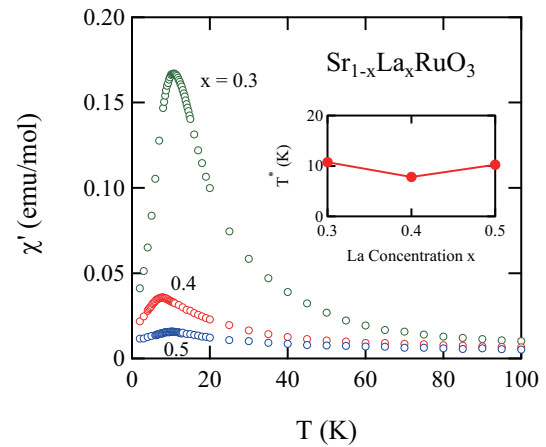}
\end{center}
\caption{(Color online) Temperature dependence of the in-phase component of ac susceptibility measured at 180 Hz for Sr$_{1-x}$La$_{x}$RuO$_3$. The inset shows the dependence of the peak temperature of  $\chi'_\mathrm{ac}$ on the La concentration.}
\label{f1}
\end{figure}

\begin{figure*}[t]
\begin{center}
\includegraphics[keepaspectratio, width=16cm,bb = 0 0 685 240,clip]{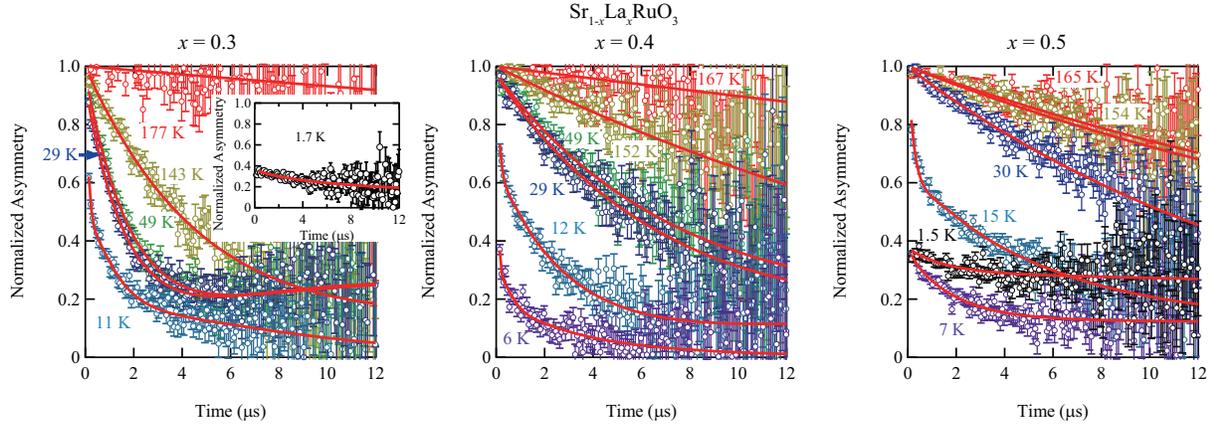}
\end{center}
\caption{(Color online)  ZF-$\mu$SR spectra at selected temperatures for Sr$_{1-x}$La$_{x}$RuO$_3$. The inset shows the ZF-spectra measured at 1.7 K for $x$ = 0.3. The background signals from the silver sample plate have been subtracted, and the spectra are normalized. The red solid lines for the spectra measured above  and below 30 K are fits using Eqs. (1) and (2), respectively. } 
\label{f1}
\end{figure*}

Figure 1 shows the temperature dependence of the dc magnetization of the present samples under field-cooled conditions with a magnetic field of 5 kOe. The dc magnetization slightly increases at 160 K, followed by further increases below approximately 50 K. The increases below approximately 50 K would be related to cluster-glass ordering, whose magnitude is markedly suppressed upon increasing $x$. The overall temperature dependence of the dc magnetization of the present sample is very similar to that obtained in our previous study.\cite{kawasaki} The increases at 160 K are extrinsic and probably caused by an impurity ferromagnetic phase of the pure SrRuO$_3$. This is because its magnitude strongly depends on the sample preparation process. The magnitude  can be reduced by iterating the sinter process, although the magnetic properties of the low-temperature cluster-glass phases depend negligibly on the sample preparation process.\cite{kawasaki} In addition, the present $\mu$SR results prove that the volume fractions are negligible, and that the magnetic fields induced by the ferromagnetic impurity phase are static and are less than 1 Oe, as we will discuss later.
The in-phase component of the ac susceptibility $\chi'_\mathrm{ac}$ of the present samples measured at 180 Hz is shown in Fig. 2.  $\chi'_\mathrm{ac}$ exhibits broad peaks near 10 K. We here define $T^{*}$ as the peak temperature of  $\chi'_\mathrm{ac}$, and the  dependence of $T^{*}$ on La concentration is displayed in the inset of Fig. 2. Although $T^{*}$ depends negligibly on  $x$, the amplitude of $\chi'_\mathrm{ac}$ at $T^{*}$ is strongly reduced upon doping La. These features and the $T^{*}$ values are comparable to those obtained in our previous study.\cite{kawasaki}

In Fig. 3, we present the ZF-$\mu$SR spectra at selected temperatures for $x$ = 0.3, 0.4, and 0.5. The spectra are normalized, and 
the signal from the silver sample plate has been subtracted. For each sample, the relaxation becomes  considerably faster below 30 K. This behavior seems to be associated with the development of magnetic clusters. For each sample, the spectra above 30 K can be well fitted by the following function:
\begin{align}
P(t) = \frac{1}{3}\mathrm{exp}(-\lambda t) + \frac{2}{3}(1-\Delta t)\mathrm{exp}(-\Delta t).
\end{align}
This functional form is based on a muon polarization function in a static Lorentzian field distribution.\cite{muon_text} An exponential term is multiplied to the first term in order to incorporate the effect of magnetic fluctuations, and the fluctuation effect on the second term is absorbed into the parameter $\Delta$. Therefore, $\lambda$ denotes the relaxation rate that originates from the fluctuations of the internal magnetic fields, and  $\Delta$ denotes the relaxation rate that is determined by the width of the static-field distribution and the magnetic fluctuations.
In the analysis of the ZF-$\mu$SR spectra, the fitting functions are broadened to account for  the width of the pulsed muon beam (68 ns). The details of the data analysis are described in Appendix. We found that the effect of the pulse width is negligible compared with the experimental uncertainty in the entire time range where the ZF-$\mu$SR spectra were measured ($t \geq$ 0.11 $\mathrm{\mu}$s$^{-1}$). A pronounced change in the ZF-$\mu$SR spectra is observed for $x$ = 0.3 above and below 160 K, which is the transition temperature of the impurity ferromagnetic phase.\cite{kawasaki} However, the initial asymmetry does not decrease across this temperature; therefore, the volume fraction of the impurity ferromagnetic phase should be negligible. The $\lambda$ value at 143 K for $x$ = 0.3 is estimated to be extremely small (1.7 $\times$ 10$^{-2}$ $\mathrm{\mu}$s$^{-1}$), which shows that the internal field is nearly static. This is because the parameter $\lambda$ reflects the magnetic fluctuation effect and is evidenced by the fact that  Eq. (1) reduces to a muon polarization function in a static Lorentzian field distribution if $\lambda$ becomes zero.\cite{muon_text}  Thus, the impurity ferromagnetic phase merely creates a static weak field distribution, whose full-width at half-maximum is about 0.4 Oe for $x$ = 0.3. The internal field due to  the impurity ferromagnetic phase becomes much smaller for $x$ = 0.4 and 0.5, as seen in Fig. 3, so the impurity ferromagnetic phase should only negligibly affect the magnetic properties of the low-temperature cluster-glass phases,  as we discussed in our previous paper.\cite{kawasaki} Above 160 K,  muon spin relaxation should be dominated by fluctuations of electronic spins and nuclear-dipole fields. For each sample, the relaxation rate  above 160 K is very small, as seen in Fig. 3, suggesting that the effect of the nuclear dipole field is also practically negligible.

For each sample, the ZF-$\mu$SR spectra below 30 K cannot be well fitted by Eq. (1) because of  the development of magnetic clusters. Therefore, we fit the ZF-$\mu$SR spectra to the following function, which assumes the presence of two components:
 \begin{align}
P(t) =&\ A_1[\frac{1}{3}\mathrm{exp}(-\lambda_1^{\beta} t^{\beta}) + \frac{2}{3}(1-\Delta_1 t)\mathrm{exp}(-\Delta_1 t)] \nonumber \\
    &+  A_2[\frac{1}{3}\mathrm{exp}(-\lambda_2 t) + \frac{2}{3}(1-\Delta_2 t)\mathrm{exp}(-\Delta_2 t)],
\end{align}
where $A_1$ and $A_2$ represent the volume fractions of the magnetically ordered and paramagnetic regions, respectively, and the total volume fraction ($A_1$ + $A_2$) is fixed to unity.  The exponent $\beta$ reflects the inhomogeneity of the magnetically ordered region,  and the ZF-$\mu$SR spectra below 30 K are satisfactorily fitted in the $\beta$ range of $0.15 \leq \beta \leq 1$, as shown in Fig. 3.  It is found that $\Delta_1$ is very large and exceeds 50 $\mathrm{\mu}$s$^{-1}$  below $T^{*}$, wheres $\Delta_2$, $\lambda_1$, and $\lambda_2$  are estimated to be several $\mathrm{\mu}$s$^{-1}$ or less in the entire temperature range below 30 K. Thus, only the second term in Eq. (2), which includes $\Delta_1$, shows a quite rapid relaxation and becomes vanishingly small in the very early time region ($t$ $\leq$ 0.2 $\mathrm{\mu}$s). This behavior allows us to estimate the volume fractions  $A_1$ and $A_2$ from the ZF-$\mu$SR spectra at around $t$ = 0.

\begin{figure}[b]
\begin{center}
\includegraphics[keepaspectratio, width=7cm,bb = 0 0 260 200,clip]{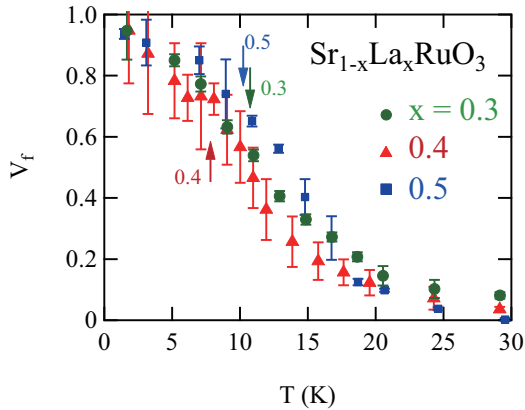}
\end{center}
\caption{(Color online) Volume fraction of magnetic ordered regions as a function of temperature for Sr$_{1-x}$La$_{x}$RuO$_3$ estimated from the ZF-$\mu$SR spectra. The arrows represent $T^{*}$. }
\label{f1}
\end{figure}

Figure 4 shows the volume fraction of the magnetically ordered region $V_f$ = $A_1$ as a function of temperature.  The positions of $T^{*}$ are indicated by arrows. We found that magnetic clusters start to develop at approximately 30 K, which is  well above $T^{*}$. Furthermore, the volume fraction increases continuously with decreasing temperature, and for each sample, it nearly reaches 100\% at the lowest temperature. The volume fraction exhibits no anomaly at $T^{*}$, which is consistent with the absence of any anomaly in the specific-heat data at  $T^{*}$.\cite{kawasaki}  Remarkably, the volume fraction shows no clear dependence on  La concentration despite the magnitude of dc magnetization being significantly suppressed with increasing $x$.  This result contrasts sharply with the Sr$_{1-x}$Ca$_x$RuO$_3$ system, where the development of the volume fraction of the magnetically ordered region is continuously suppressed with increasing Ca concentration over the entire range of Ca concentration.\cite{imgat}

As described above,  $\Delta_1$ is significantly larger than the other parameters; hence, $\Delta_1/\gamma_\mu$ mainly represents the distribution width of the local field in the magnetically ordered region. Here, $\gamma_\mu$ is the muon gyromagnetic ratio. However, in the present study, we cannot correctly estimate $\Delta_1$  because  it becomes too large, especially below  $T^{*}$; therefore, the second term in Eq. (2), which includes $\Delta_1$, becomes vanishingly small before initiating the data acquisition. On the other hand, because the volume fraction  in each sample of the magnetically ordered region is nearly 100\% at the lowest temperature, the magnitude of dc magnetization is expected to reflect the internal field of the magnetic clusters; thus, the internal field appears to be significantly suppressed upon increasing $x$.

\begin{figure}[h]
\begin{center}
\includegraphics[keepaspectratio, width=7cm,bb = 0 0 260 200,clip]{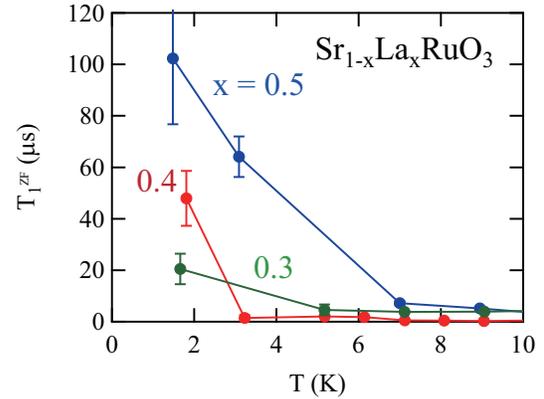}
\end{center}
\caption{(Color online) Relaxation time [defined by Eq. (3)] of the ZF-$\mu$SR spectra as a function of temperature. }
\label{f1}
\end{figure}

We now focus on the dynamics of magnetic clusters.  
Below $T^{*}$, the ZF-$\mu$SR spectra are mainly dominated by the first term in Eq. (2), $\frac{1}{3}$exp$(-\lambda_1^{\beta} t^{\beta})$, which describes the relaxation process caused by magnetic fluctuations in the magnetically ordered region. Therefore, the functional form is mainly determined by only two fitting parameters, and we can safely estimate these parameters at low temperatures below $T^{*}$. 
Here, we introduce the relaxation time  of muon spins as 
\begin{align}
T_1^\mathrm{ZF} = \frac{[\mathrm{ln}(2)]^{1/\beta}}{\lambda_1}.
\end{align}
This relaxation time is defined as the time at which the first term in Eq. (2) reduces to half of its initial value and allows us to compare the magnetic fluctuation effects of the spectra fitted with different $\beta$ values. The temperature dependence of  $T_1^\mathrm{ZF}$ is shown in Fig. 5. The quantity $T_1^\mathrm{ZF}$ increases with decreasing temperature and exceeds the time range of $\mu$SR experiments at the lowest temperatures, which implies that at the lowest temperatures, magnetic fluctuations caused by magnetic clusters completely disappear on the time scale of $\mu$SR experiments.

\begin{figure}[h]
\begin{center}
\includegraphics[keepaspectratio, width=7cm,bb = 0 0 270 450,clip]{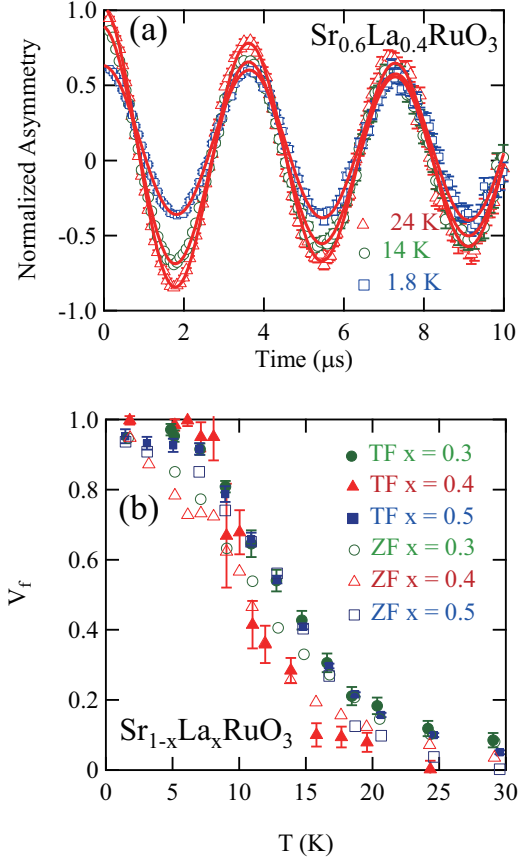}
\end{center}
\caption{(Color online) (a)  TF-$\mu$SR spectra at selected temperatures for $x$ = 0.4. The spectra are normalized. The red solid curves are fits to Eq. (4).  (b)  Temperature variation of the volume fraction of the magnetically ordered region estimated from the TF-$\mu$SR spectra. The volume fraction estimated from the ZF-$\mu$SR spectra is also shown.}
\label{f1}
\end{figure}

We  also performed $\mu$SR measurements under weak transverse fields (WTFs), which provide an alternative method to determine  the volume fraction of the magnetically ordered region. In WTF measurements, muon spins in the paramagnetic region precess around the applied WTF at the Larmor frequency, whereas muon spins in  the magnetically ordered regions yield no precession signal but instead exhibit a fast depolarization due to the strong internal fields. Therefore, the volume fraction of the nonmagnetic region can be estimated from the amplitude of the precession signal. Figure 6(a) shows the $\mu$SR spectra under a WTF of 20 Oe at selected temperatures for $x$ = 0.4. With decreasing temperature, the amplitude of the precession signal decreases, which indicates the development of magnetic clusters. We analyze the $\mu$SR spectra under WTFs using the following function:
\begin{align}
P(t) =&\ A_1\mathrm{exp}(-\lambda_1 t) + A_2\mathrm{cos}(\omega t+\phi)\mathrm{exp}(-\lambda_2 t) \nonumber \\
     &+ A_\mathrm{Ag}\mathrm{cos}(\omega t+\phi)\mathrm{exp}(-\sigma_\mathrm{Ag}^2 t^2), 
\end{align}
where the first and second terms represent the magnetically ordered and paramagnetic components, respectively. The third term is the contribution from the silver sample plate. 
Figure 6(b) displays the temperature dependence of the volume fraction of the magnetically ordered region (1 - $A_2$ - $A_\mathrm{Ag}$) /(1 - $A_\mathrm{Ag}$). Here, we also plot the volume fraction estimated from the ZF-$\mu$SR spectra (open symbols).  The volume fractions estimated from the WTF measurements are almost the same as those estimated from the ZF-$\mu$SR spectra. Hence, this result fully supports the validity of our analysis performed on the ZF-$\mu$SR spectra.

\begin{figure}[h]
\begin{center}
\includegraphics[keepaspectratio, width=7cm,bb = 0 0 260 200,clip]{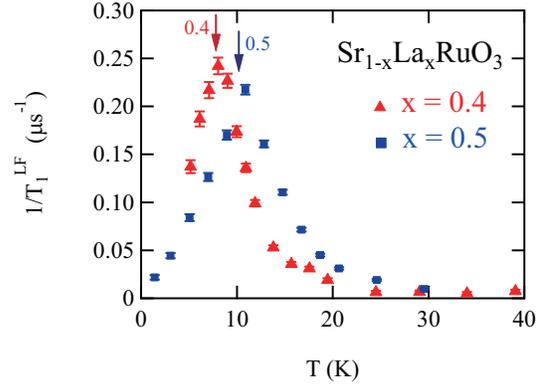}
\end{center}
\caption{(Color online)  LF-relaxation rates at an applied field of 200 Oe as a function of temperature. The positions of  $T^{*}$ are indicated by arrows. }
\label{f1}
\end{figure}

In order to gain further insight into the magnetic fluctuations of this system, we carried out $\mu$SR experiments at a LF of 200 Oe for $x$ = 0.4 and 0.5. The LF-spectra  (not shown) can be fitted by a single exponential function $A_1 \mathrm{exp}(-t/T_1^\mathrm{LF})$, where $1/T_1^\mathrm{LF}$ is the relaxation rate. Note that this relaxation rate differs from the relaxation rate $1/T_1^\mathrm{ZF}$ estimated from the ZF-$\mu$SR spectra: the former represents the effect of magnetic fluctuations over the entire sample whereas the latter reflects the effect of magnetic fluctuations only in the magnetically ordered region. In Fig. 7, we plot the relaxation rate  $1/T_1^\mathrm{LF}$ as a function of temperature, with the positions of $T^{*}$ indicated by arrows. $1/T_1^\mathrm{LF}$ exhibits divergent-like behavior near $T^{*}$, implying that the magnetic fluctuations are considered to have a critical character near $T^{*}$. Thus, the  $\mu$SR data show that  the peak in $\chi'_\mathrm{ac}$ is related to the development of  magnetic fluctuations rather than to the increase in the volume fraction of the magnetically ordered region.

\begin{figure}[t]
\begin{center}
\includegraphics[keepaspectratio, width=6cm,bb = 0 0 270 670,clip]{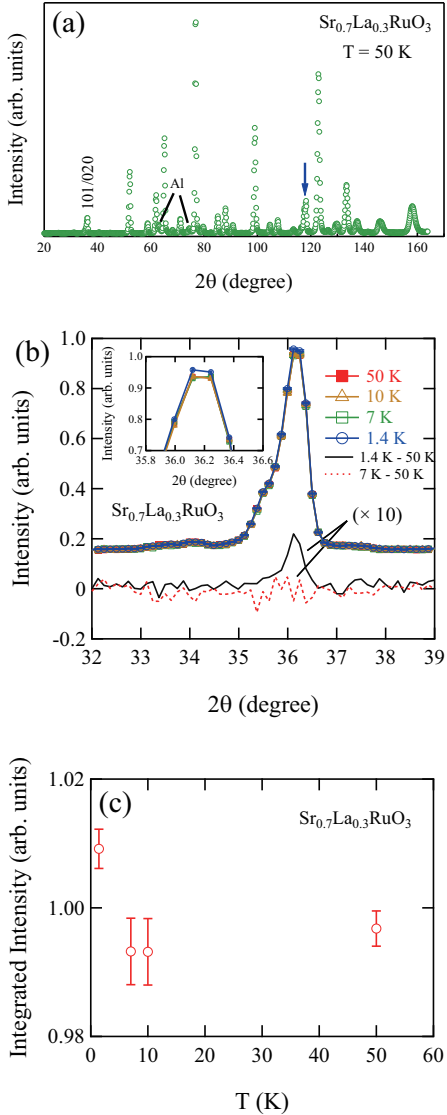}
\end{center}
\caption{(Color online) (a) Neutron diffraction pattern of Sr$_{0.7}$La$_{0.3}$RuO$_3$ obtained at 50 K. 
The blue arrow shows the position of the unknown peak. (b) Temperature variation of the (101) and (020) reflections. The differences between low temperatures and 50 K are shown as the black solid and red broken lines, which are ten times enlarged.  The inset shows  an enlarged view of the spectra around the peak position. (c) Temperature dependence of integrated intensity of the (101) and (020) reflections.}
\label{f1}
\end{figure}

To further characterize the cluster-glass state in Sr$_{1-x}$La$_x$RuO$_3$, we performed neutron powder diffraction experiments for $x = 0.3$ over the temperature range of 1.4-50 K. The sample used in this neutron experiment is different from those used in the  ac susceptibility, dc magnetization, and $\mu$SR experiments, and $T^{*}$ of this sample is about 14 K. Figure 8(a) shows a neutron powder diffraction pattern measured at 50 K, where the volume fraction of the magnetically ordered region is negligible according to the $\mu$SR results. Hence, the contribution of the magnetic scattering can be neglected.  All  the observed diffraction peaks [except for the peaks attributed to aluminum and an unknown peak indicated by the blue arrow in Fig. 8(a)] are indexed to the space group $Pnma$.

Figure 8(b) shows the temperature dependence of the (101) and (020) reflections, which are expected to have the largest magnetic contributions associated with ferromagnetic order. The integrated intensity of these reflections as a function of temperature is displayed in Fig. 8(c). The enhancement in the integrated intensity appears only at the lowest temperature (1.4 K), indicating the development of  ferromagnetic order. Note that no clear increase is observed down to 7 K, although this temperature is less than $T^{*}$, and the volume fraction of the magnetically ordered region is approximately 80\% at this temperature. As seen in the difference plot between the 1.4 and 50 K neutron powder diffraction patterns, the peak width of the magnetic reflection is comparable to the nuclear peak width; thus, within the resolution of our measurements, the ferromagnetic order detected at the lowest temperature is of long range ($\geq$1000 \AA).

The ferromagnetic ordered moment must belong to a single irreducible representation of the point group of this system $D_{2h}$.\cite{group} Therefore, the possible directions of the magnetic moments are restricted to [100], [010], and [001]. 
However, we cannot determine the direction of the magnetic moment in this study because the magnetic reflections are too small and  because this system has a nearly cubic structure, which results in overlapping of several diffraction peaks.\cite{group} Using the magnetic form factor of Sr$_2$RuO$_4$, we estimate the magnitude of the ordered moment at 1.4 K to be approximately 0.5 $\mu_\mathrm{B}$.\cite{nagata}

\section{Discussion}

In the previous and present studies, we performed  the ac susceptibility, dc magnetization, $\mu$SR, and neutron powder diffraction measurements on the cluster-glass states in  Sr$_{1-x}$La$_x$RuO$_3$.\cite{kawasaki} We obtained the following results: (1)  $T^{*}$ varies clearly with frequency, and the initial frequency shift $\delta$ =  $\Delta T^{*}/(T^{*}\Delta \mathrm{log_{10}}\omega$) is estimated to range from 0.040 to 0.084 for 0.3 $\leq$ $x$  $\leq$ 0.5. This result for $\delta$ is comparable to that of cluster-glass systems. We also found that the frequency dependence is well described by the Vogel-Fulcher law.\cite{vogel,fulcher}  (2) Magnetic clusters start to develop well above $T^{*}$, and the volume fraction of the magnetically ordered region gradually increases with decreasing temperature without showing any anomaly at $T^{*}$ and reaches nearly 100\% at the lowest temperature. This development is expected to be responsible for the increase in the dc magnetization at low temperatures. (3) No clear dependence of La concentration  on the temperature variation of volume fractions is observed over the presently investigated $x$ range; however, the magnitude of the dc magnetization and ac susceptibility is markedly suppressed upon increasing $x$. (4)  No clear magnetic scattering is observed in the neutron powder diffraction experiments for $x$ = 0.3 down to 7 K, which is below $T^{*}$, and a long-range ferromagnetic order  is detected at the lowest temperature. The results (2) and (4) obtained in this study are consistent with the cluster-glass ordering,  as discussed below, further supporting the presence of the cluster-glass state in this system.

The results (1) and (2) clearly establish that the magnetic ordering process for $x$ $\geq$ 0.3 differs significantly from that expected for a conventional second-order ferromagnetic transition, and that the gradual development of the volume fraction of the magnetically ordered region  is consistent with the cluster-glass behavior. Therefore,  $T^{*}$ is no longer a real phase transition but rather a gradual freezing. Thus, although critical-like magnetic fluctuations are detected by the LF-$\mu$SR experiments near $T^{*}$, they cannot simply be understood within the framework of the SCR theory for itinerant ferromagnets.\cite{moriya}.

In the previous paper,\cite{kawasaki} we  discussed the properties of the cluster-glass states in  Sr$_{1-x}$La$_x$RuO$_3$ based on the optimal fluctuation theory combined with a finite-size scaling technique.\cite{hrahsheh} This theory assumes that the magnetic property of a local region is characterized by the local La concentration and the size of the local region. If we assume a random distribution of Sr and La, the probability that a small region contains  $N_\mathrm{La}$ number of La atoms can be calculated using the binomial distribution $P(N_\mathrm{La})=\binom{N}{N_\mathrm{La}}x^{N_\mathrm{La}}(1-x)^{N-N_\mathrm{La}}$. Furthermore, it is assumed that the small region has magnetic order when its local La concentration $x_\mathrm{loc}=N_\mathrm{La}/N$ is less than the threshold $x_\mathrm{c}(L_\mathrm{RR})$. The functional form of  $x_\mathrm{c}(L_\mathrm{RR})$ is given by $x_\mathrm{c}(L_\mathrm{RR})=x_\mathrm{c}^0 - DL_\mathrm{RR}^{-\phi}$ based on the finite size scaling  argument, where $x_\mathrm{c}^0$ is the critical concentration for the bulk system, $D$ is a constant, and $\phi$ is the finite-size shift exponent. This theory applies to the Sr$_{1-x}$Ca$_x$RuO$_3$ system and successfully explains the dependence of the Ca concentration on the low-temperature dc magnetization data.\cite{demko}  In Sr$_{1-x}$La$_x$RuO$_3$, however, the result (3) clearly contradicts this assumption, suggesting that the properties of cluster-glass states in Sr$_{1-x}$La$_x$RuO$_3$ cannot be explained simply in terms of the distribution of La atoms in real space.

Interestingly, for $x$ = 0.3, long-range ferromagnetic order is detected  by neutron powder diffraction measurements at 1.4 K. Here, we discuss the implications of this result. According to the $\mu$SR results, the magnetic clusters at $x$ = 0.3 start to develop at approximately 30 K and are considered to be dynamically fluctuating  down to  $T^{*}$. Below $T^{*}$, the magnetic clusters seem to freeze  in the time scale of the ac susceptibility measurement.\cite{kawasaki} The strong dependence $T^{*}$ on frequency detected by the ac susceptibility measurements clearly suggests that the ordered states around $T^{*}$ is a cluster-glass state, which consists of short-range  ferromagnetic clusters. Since no anomaly appears in the  ac susceptibility and dc magnetization data below $T^{*}$, the observation of long-range ferromagnetic order at the lowest temperature implies that the magnetic correlation length increases with decreasing temperature, and that the cluster-glass state gradually changes to a long-range ferromagnetic ordered state.  The presence of the ferromagnetic ground state is consistent with the rapid increase in  $T_1^\mathrm{ZF}$ at the lowest temperatures.  Thus, the absence of magnetic contribution in the neutron powder diffraction pattern at 7 K, which is below $T^{*}$, may possibly be due to the magnetic correlation length being too short to be detected by neutron diffraction experiments at this temperature. A similar crossover was reported for the CeNi$_{1-x}$Cu$_x$ system.\cite{marcano}  For CeNi$_{1-x}$Cu$_x$, the cluster-glass order is revealed by ac susceptibility measurements as in Sr$_{1-x}$La$_x$RuO$_3$, and neutron diffraction studies showed that the cluster-glass state continuously changes into a long-range ferromagnetic state. However,  the temperature dependence of the volume fraction of magnetic clusters for CeNi$_{1-x}$Cu$_x$ differs from that of Sr$_{1-x}$La$_x$RuO$_3$ and already reaches 100\% at   $T^{*}$.

\section{Conclusions}

The cluster-glass states of Sr$_{1-x}$La$_x$RuO$_3$ (0.3 $\leq$ $x$ $\leq$ 0.5) were studied by $\mu$SR and neutron powder diffraction measurements. We found from the ZF- and TF-$\mu$SR experiments that the volume fraction of the magnetically ordered region starts to develop well above $T^{*}$, and that it increases continuously with decreasing temperature reaching nearly 100\% at the lowest temperature, with no anomaly at $T^{*}$. Although the dc magnetization is strongly suppressed with increasing $x$, no clear dependence of $x$ on the temperature variation of the volume fraction of the magnetic clusters is observed over the presently investigated $x$ range.  The LF-$\mu$SR experiments have shown that $1/T_1^\mathrm{LF}$ exhibits divergent-like behavior near $T^{*}$, indicating a critical character of magnetic fluctuations. Long-range ferromagnetic order is detected by neutron powder diffraction experiments for $x$ = 0.3 at the lowest temperature, whereas cluster-glass ordering below $T^{*}$ is suggested by the previous ac susceptibility and dc magnetization studies. The absence of any anomaly in ac susceptibility and dc magnetization data below $T^{*}$ implies that  cluster-glass states in Sr$_{1-x}$La$_{x}$RuO$_3$ gradually develop with decreasing temperature into long-range ferromagnetic ordered states.   These results suggest that the magnetic ordering process of the cluster-glass phases in Sr$_{1-x}$La$_x$RuO$_3$ differs strikingly from that expected for a conventional second-order ferromagnetic transition.

\begin{acknowledgments}
This work was supported by a Grant-in-Aid for Young Scientists (B) (Grant No. 25800211) from the Ministry of Education, Culture, Sports, Science and Technology of Japan.  The neutron diffraction experiment was conducted with Echidna at the OPAL facility of the Australian Nuclear Science and Technology Organisation, using beam time that was transferred from HERMES at JRR-3 with the approval of ISSP, The University of Tokyo, and JAEA.

\end{acknowledgments}

\appendix
\section{Analysis of $\mu$SR Spectra Considering Pulse Width Effect}
The muon beam at the RIKEN-RAL Muon Facility has a pulsed time structure determined by the time structure of the primary proton beam. The width of a single muon pulse is 68 ns.  Since the limited time resolution caused by the pulse width may affect $\mu$SR spectra, it is necessary to take into account its effect in the analysis in order to  obtain reliable fitting parameters.   
 
Here, we derive the  asymmetry of the pulsed $\mu$SR $A^P(t)$ from the asymmetry of the conventional $\mu$SR $A^0(t)$. The positron counts  at the forward and backward counters $N_{+}(t)$ and $N_{-}(t)$ in the conventional $\mu$SR experiment  can be written as 
\begin{align}
N^0_\pm(t)=N^0_\pm\mathrm{exp}(-t/\tau_{\mu})[1\pm A^0(t)], 
\end{align}
where $\tau_{\mu}$ = 2.2 $\mu$s is the muon lifetime. Therefore,  $A^0(t)$ is given by
\begin{align}
A^0(t)=&\ \frac{N^0_{+}(t)-\alpha N^0_{-}(t)}{N^0_{+}(t)+\alpha N^0_{-}(t)}, \\
     \alpha=&\ N^0_{+}/N^0_{-}. 
\end{align}
We can derive the expression for $A^P(t)$ from Eq. (A.2) by incorporating the time structure  $b_\mu(t)$ of the muon beam:
\begin{align}
A^P(t)=&\ \frac{\int N^0_{+}(t+\delta)b_\mu(\delta)d\delta-\alpha\int N^0_{-}(t+\delta)b_\mu(\delta)d\delta}{\int N^0_{+}(t+\delta)b_\mu(\delta)d\delta+\alpha\int N^0_{-}(t+\delta)b_\mu(\delta)d\delta} \nonumber \\
 =&\ \frac{\int e^{-\delta/t}A^0(t + \delta)b_\mu(\delta)d\delta}{\int e^{-\delta/t}b_\mu(\delta)d\delta}.
\end{align}
 In the same way, the broadened model function $G^P(t)$ for fitting  the pulsed $\mu$SR spectra can be obtained from that for conventional  $\mu$SR spectra $G^0(t)$: 
\begin{align}
G^P(t)=\frac{\int e^{-\delta/t}G^0(t + \delta)b_\mu(\delta)d\delta}{\int e^{-\delta/t}b_\mu(\delta)d\delta}.
\end{align}
In the present study, we approximate  the time structure of the muon beam $b_\mu(t)$ by a Gaussian function with a full width at half-maximum of 68 ns. We analyzed the ZF-$\mu$SR spectra using the above functional form and found that  the effect of pulse width is negligible in the present analysis.

\end{document}